\documentclass[11pt]{article}
\usepackage{xspace}
\usepackage{graphicx}
\usepackage{amsmath}
\usepackage{amssymb}
\usepackage{color}
\usepackage{caption}
\usepackage{subfigure}

\definecolor{Red}{rgb}{1,0,0}
\definecolor{Green}{rgb}{0,1,0}
\definecolor{Blue}{rgb}{0,0,1}
\definecolor{Black}{rgb}{0,0,0}

\def\beq{\begin{equation}}
\def\eeq#1{\label{#1}\end{equation}}
\def\eeqn{\end{equation}}

\def\beqa{\begin{eqnarray}}
\def\eeqa#1{\label{#1}\end{eqnarray}}
\def\eeqan{\end{eqnarray}}

\let\bar=\overbar

\def\Dslash{\not{\hbox{\kern-4pt $D$}}}
\def\dslash{\not{\hbox{\kern-2pt $\del$}}}

\def\msb{{\bar{\ssstyle M \kern -1pt S}}}

\def\Title#1{\begin{center} {\Large {\bf #1} } \end{center}}

\newcommand{\mev}{\ \mbox{MeV}}

\newcommand{\nue}{\nu_e}
\newcommand{\nuebar}{\bar{\nu_e}}
\newcommand{\numu}{\nu_\mu}
\newcommand{\numubar}{\bar{\nu_\mu}}

\begin{document}

\Title{Measuring the electron anti-neutrino beam component in the T2K near detector ND280}

\bigskip\bigskip

%+\addtocontents{toc}{{\it D. Reggiano}}
%+\label{ReggianoStart}

\begin{raggedright}  

%% Authors - you should specify at least one author as follows.
{\it Luke Southwell\index{Southwell, L.},\\
Department of Physics\\
Lancaster University\\
LA1 4YW Lancaster, UK}\\
%% In case you want to have more than one author please follow the format
%% shown below, listing the individual authors AND also making sure
%% that each author is given a unique index entry.
%Someone Else\index{Else, S.}, {\it Another University}\\

\end{raggedright}
\vspace{1.cm}

{\small
\begin{flushleft}
\emph{To appear in the proceedings of the Prospects in Neutrino Physics Conference, 15 -- 17 December, 2014, held at Queen Mary University of London, UK.}
\end{flushleft}
}

\begin{abstract}
The main irreducible background in the T2K $\nue$ appearance analysis is the $\nue$ contamination in the $\numu$ beam. In order to quantify this background, a selection for charged-current $\nue$ interactions in the near detector (ND280) tracker region was developed by combining the particle identification abilities of the time projection chambers and electromagnetic calorimeters. We measured a data/MC ratio of $1.01 \pm 0.10$ for the $\nue$ component of the beam which, is an important confirmation of our predictions of the expected backgrounds. In 2014 the T2K experiment reversed the polarity of the magnetic horns and began running with an antineutrino beam for the first time. Differences in the oscillation probabilities between neutrinos and antineutrinos may provide insight into CP violation in the leptonic sector. The current ND280 Tracker $\nue$ charged-current selection has been used as a starting point for the $\nuebar$ charged-current selection. The additional challenges and selection criteria of the electron anti-neutrino selection will be presented.

\end{abstract}

\section{The T2K Experiment}

The T2K experiment is a long-baseline neutrino oscillation experiment which uses a $\numu$ beam. The objective of the experiment is to measure the oscillation parameter $\theta_{13}$ via $\nue$ appearance and the parameters $\Delta m^{2}_{32}$ and $\theta_{23}$ via $\numu$ disappearance\cite{T2K}. Neutrino interactions are observed in the T2K far detector Super-Kamiokande (SK) which is 295 km away from the beam source and 2.5 degrees off axis\cite{T2K}. The $\numu$ beam also contains contamination from $\nue$, $\numubar$ and $\nuebar$.

Neutrinos are also observed with the on-axis and off-axis near detectors 280 m from the beam source. The off-axis detector ND280, is used to measure neutrino interaction properties and the contamination in the beam from other flavours of neutrinos. It has several sub-detectors: an upstream $\pi^0$ detector (P0D) followed by a tracker region comprising three gaseous argon time projection chambers (TPCs) interspersed with two scintillator-based fine grained detectors (FGD1 and FGD2). The P0D and tracker region are surrounded by a set of electromagnetic sampling calorimeters (ECals) consisting of alternating layers of lead and scintillator. The yoke of the magnet is also instrumented with plastic scintillator to form side muon range detectors (SMRDs)\cite{T2K}.

\section{Selecting $\nue$ Charged-Current Events in the ND280 Tracker}

To select $\nue$ CC inclusive interactions in the tracker region, FGD1 and FGD2 are used as the target mass. Events in which there are electron-like tracks are selected using TPC particle identification (PID) criteria that are based on the rate of energy loss as the particle traverses the detector (dE/dx).  Following the application of PID criteria, the sample is 92$\%$ pure in electrons but only 27$\%$ of these electrons originate from a $\nue$ interaction, with the majority of the non-$\nue$ events originating from photons converting to an $e^+e^-$ pair in the FGDs. To reduce this contamination, an upstream veto is applied which rejects events with tracks in the P0D, TPC, or ECals that start upstream from the initial position of the electron candidate. If an electron-like positive track is within 10 cm of the electron candidate and the pair of tracks have a reconstructed mass of less than 100 $\mev$/$\mbox{c}^{2}$, the event is rejected. Following these cuts the contamination is reduced from 65$\%$ to 30$\%$. A more detailed description can be seen in reference \cite{bensmith}.

\begin{figure}[!ht]
   \centering
   \subfigure[]{\label{fig:3}\includegraphics[width=0.4\textwidth]{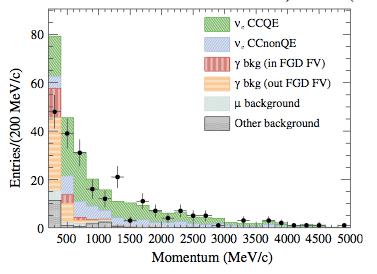}}
   \subfigure[]{\label{fig:4}\includegraphics[width=0.4\textwidth]{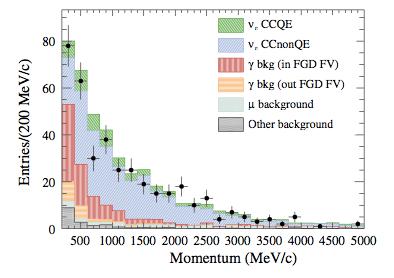}}
\caption{Reconstructed electron momentum of events in the CCQE-like sample (\ref{fig:3}) and in the CCnonQE-like sample (\ref{fig:4}). The errors on the data points are statistical. The coloured histogram is Monte Carlo. The CCQE events are shown in green and the CCnonQE events are in blue. The dominant background from gamma conversions being shown in red or orange, differentiating if the neutrino which produced the photon interacted inside the FGD fiducial volume or outside it. The other background is mainly misidentified pions \cite{bensmith}.}
\end{figure}

Further criteria are then applied to separate the $\nue$ CC inclusive sample into a CC quasi-elastic sample (CCQE) (figure \ref{fig:3}) and a CC non-quasi-elastic sample (CCnonQE) (figure \ref{fig:4}). The CCQE sample is 48$\%$ pure with an efficiency of 36.5$\%$ and the CCnonQE sample is 53$\%$ pure with an efficiency of 30.6$\%$.

For the CC inclusive sample the ratio of data over Monte Carlo is $1.01 \pm 0.10$. A further measurement was performed by independently fitting $\nue$ originating from $\mu^+$ and kaon parents using the CC inclusive sample. The data over Monte Carlo ratios were found to be $0.68 \pm 0.30$ for $\nue$ originating from $\mu^+$ and $1.10 \pm 0.14$ for $\nue$ originating from kaons \cite{bensmith}. Overall the measured $\nue$ contamination in the T2K beam is $(1.2 \pm 0.1)$$\%$ \cite{bensmith}.

\section{Selecting $\nuebar$}

\begin{figure}[!ht]
   \centering
   \subfigure[]{\label{fig:5}\includegraphics[width=0.4\textwidth]{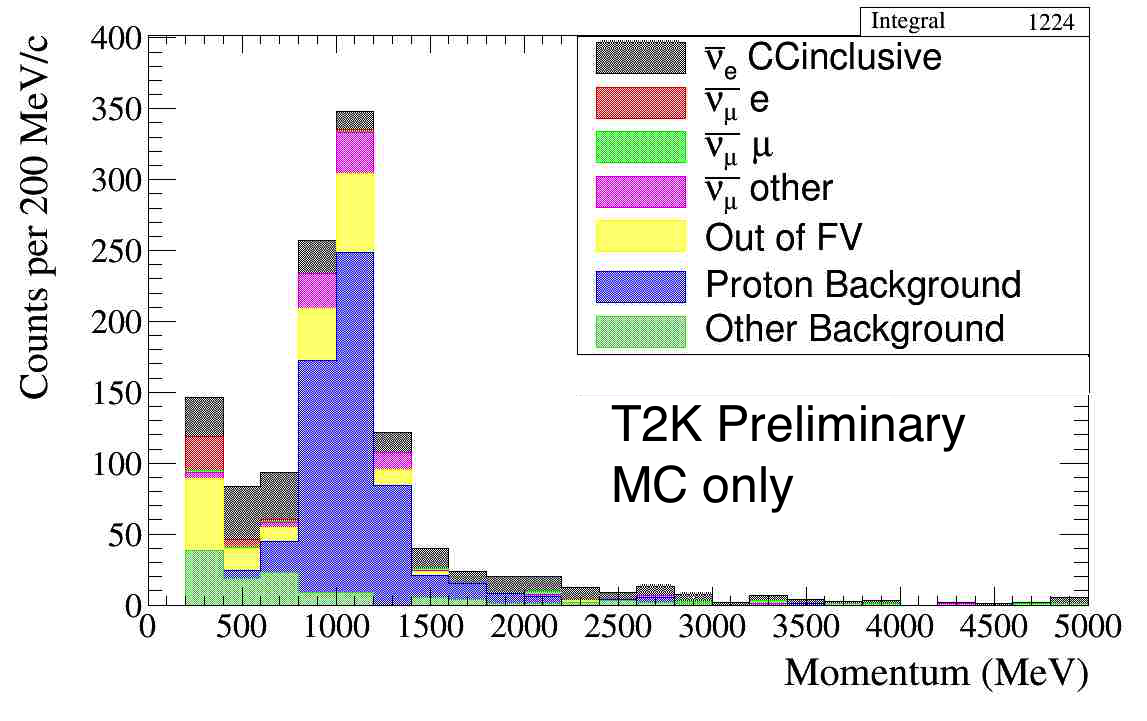}}
   \subfigure[]{\label{fig:6}\includegraphics[width=0.4\textwidth]{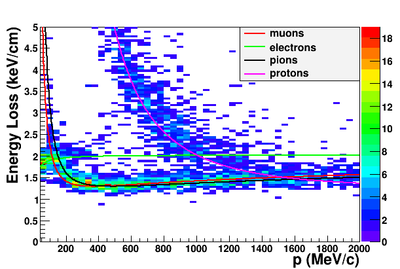}}
\caption{The $\nuebar$ sample after only changing the charge requirement (\ref{fig:5}). The large proton background (blue) can be seen in the area around 1000 $\mev$/$\mbox{c}$ dominating over the $\nuebar$ signal (black). It can be seen in figure \ref{fig:6} that the proton and positron curves have similar values around 1000 $\mev$/$\mbox{c}$ momentum.}
\end{figure}

The $\nuebar$ contamination in the T2K anti-neutrino beam has not yet been measured. In order to create a $\nuebar$ selection, the existing T2K $\nue$ selection was modified by reversing the charge requirement to look for positive particles instead of negative ones. However, this immediately presented new challenges: protons, which were previously removed by the negative charge requirement, now pass the selection cuts. This background can be seen in figure \ref{fig:5}, which shows the MC prediction for the $\nuebar$ sample.

\begin{figure}[!ht]
\centering
\includegraphics[width=.4\textwidth]{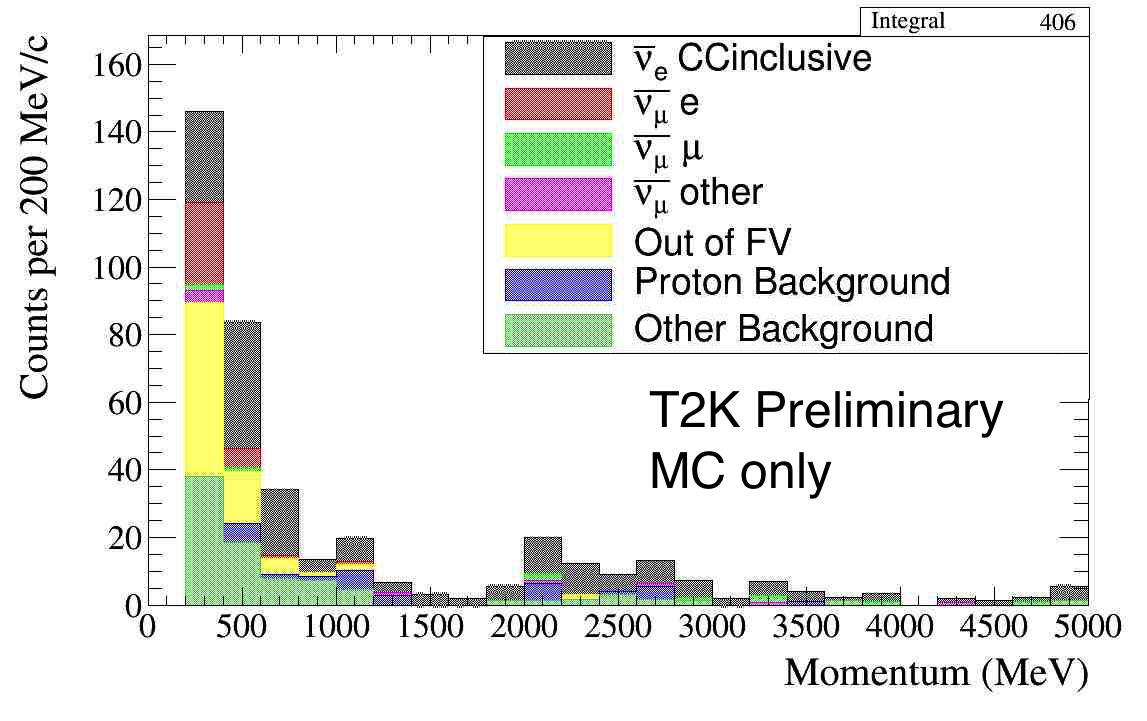}
\caption{The $\nuebar$ sample after the full selection has been applied. The $\nuebar$ signal can be seen in black and the rest of the events are predominantly gamma background as in the $\nue$ selection. The other background is also predominantly gamma. }
\label{fig7}
\end{figure}

As mentioned above, the TPC PID depends on $dE/dx$. As can be seen in figure \ref{fig:6}, the positron and proton dE/dx are similar around a momentum of 1000$\mev$/$\mbox{c}$; therefore, the TPC PID cannot be used to discriminate between protons and positrons in this region. In order to remove the proton background, the ratio of electromagnetic energy deposited in the ECal to the momentum measured by the TPCs ($E/p$) was used, in the region where the proton background dominates. For a given momentum, the proton has far less kinetic energy to deposit in the ECals than a positron, so $E/p$ is higher for positrons. Additional ECal PID variables, combining the number of ECal hits and the patterns of energy deposition in the ECal into log-likelihood variables, are also used for distinguishing between particles to further improve the selection. This results in a 96$\%$ reduction of the proton background as seen in figure \ref{fig7}.

\section{$\nuebar$ results}

The criteria described in section 3 result in a $\nuebar$ selection with a purity of $(42.4 \pm 3.6)$$\%$ and an efficiency of $(32.7 \pm 2.8)$$\%$  for $\nuebar$ CC interactions in the FGD fiducial volume. This selection will be applied to the T2K antineutrino beam data. The uncertainties are MC statistical only.

\section{Summary}
The $\nuebar$ selection is now complete with a purity of $(42.4 \pm 3.6)$$\%$ and an efficiency of $(32.7 \pm 2.8)$$\%$. Systematic studies are ongoing. Preliminary results indicate that systematic uncertainties will be slightly larger than those presented in \cite{bensmith} due to the uncertainty associated with the proton background. Once the systematics have been fully analysed we will be ready to use the selection on antineutrino data taken by the T2K experiment to measure the $\nuebar$ contamination in the T2K anti-neutrino beam, which is expected to be of order 1.0$\%$. This contamination feeds into the T2K $\nuebar$ appearance oscillation measurement as it will be the main irreducible background.

\bigskip
\section{Acknowledgments}

This work was presented on behalf of the T2K collaboration. The presenter would like to thank Dr. Laura Kormos and the T2K publication board for their input.

\end{document}